\pdfoutput=1
\documentclass{article}
\usepackage{spconf,amsmath,graphicx}
\usepackage{amssymb}
\usepackage{setspace}
\usepackage{multirow}
\usepackage{cite}
\usepackage{url}
\usepackage{graphicx}
\usepackage{bbding}
\usepackage{enumerate}

\title{PAVITS: EXPLORING PROSODY-AWARE VITS FOR END-TO-END EMOTIONAL VOICE CONVERSION}
\name{Tianhua~Qi$^{1,2}$, Wenming~Zheng$^{1,2*}$, Cheng~Lu$^{1,2}$, Yuan~Zong$^{1,2}$, Hailun~Lian$^{1}$ 
\thanks{This work was supported in part by the National Key R \& D Project under the Grant 2022YFC2405600, in part by the NSFC under the Grant U2003207 and 61921004, and in part by the Jiangsu Frontier Technology Basic Research Project under the Grant BK20192004.}
}
\address{
	$^1$Key Laboratory of Child Development and Learning Science (Southeast University),\\ Ministry of Education, Nanjing 210096, China\\
	$^2$School of Biological Science and Medical Engineering, Southeast University, China\\
	\{qitianhua, wenming\_zheng$^{*}$, cheng.lu, xhzongyuan, lianhailun\}@seu.edu.cn
}

\begin{document}
\ninept
\maketitle{\tiny }
\begin{abstract}
In this paper, we propose Prosody-aware VITS (PAVITS) for emotional voice conversion (EVC), aiming to achieve two major objectives of EVC: high content naturalness and high emotional naturalness, which are crucial for meeting the demands of human perception. To improve the content naturalness of converted audio, we have developed an end-to-end EVC architecture inspired by the high audio quality of VITS. By seamlessly integrating an acoustic converter and vocoder, we effectively address the common issue of mismatch between emotional prosody training and run-time conversion that is prevalent in existing EVC models. To further enhance the emotional naturalness, we introduce an emotion descriptor to model the subtle prosody variations of different speech emotions. Additionally, we propose a prosody predictor, which predicts prosody features from text based on the provided emotion label. Notably, we introduce a prosody alignment loss to establish a connection between latent prosody features from two distinct modalities, ensuring effective training. Experimental results show that the performance of PAVITS is superior to the state-of-the-art EVC methods. Speech Samples are available at \url{https://jeremychee4.github.io/pavits4EVC/}.
\end{abstract}
\begin{keywords}
Emotional voice conversion, end-to-end model, prosody, emotional speech, multi-task learning
\end{keywords}
\section{Introduction}
\label{sec:intro}

Emotional voice conversion (EVC) endeavors to transform the state of a spoken utterance from one emotion to another, while preserving the linguistic content and speaker identity~\cite{Zhou2023}. It brings the capability to facilitate emotional communication between individuals~\cite{lu2021multi}, enhancing the user experience in human-computer interaction~\cite{chatterjee2021real}, and even achieving a seamless integration of human presence within the virtual world~\cite{dionisio20133d}.

There are two distinct challenges in EVC: one is low content naturalness, and the other is that the converted audio lacks the richness of emotion compared to human voice\cite{Zhou2023}. Previous studies were focused on frame-based solutions, such as CycleGAN~\cite{fu2022improved} and StarGAN~\cite{du2021expressive, he2021improved}. However, due to the fixed-length nature and poor training stability, the naturalness of converted audio is quite low to apply in practice. To address this challenge, autoencoder-based~\cite{chen2022speaker, lu2022one} especially for sequence-to-sequence (seq2seq)~\cite{yang2022overview, zhao2021improving} frameworks raise much interests for its variable-length speech generation. It achieves an acceptable naturalness through the joint training with Text-to-speech (TTS)~\cite{kim2020emotional}, which is used to capture linguistic information and avoid mispronunciation as well as skipping-words. 
Since speech emotion is inherently supra-segmental~\cite{rajamani2021novel}, it is difficult to learn emotional representation from the spectrogram. To tackle this, various pretraining methods, such as leveraging speech emotion recognition (SER) model~\cite{zhou2021seen} and 2-stage training strategy~\cite{zhou2021limited}, are introduced to extract emotional feature for EVC system.

Despite these works have achieved great success in EVC, the converted audio still falls short in meeting human's perceptual needs, which implies that these two challenges still remain to be effectively addressed. Remarkably, current EVC models generally operate in a cascade manner, i.e., the acoustic converter and the vocoder~\cite{fu2022improved, he2021improved,chen2022speaker, Zhou2023}, resulting in a mismatch between emotional prosody training and run-time conversion, ultimately leading to a degradation in audio quality, which is vital to evaluate content naturalness and impacts the perceptual experience of emotional utterance. However, there is no EVC model that attempt to bridge this gap, let alone models that aim to capture prosody variations at a finer granularity. To handle the similar issue, multiple solutions have been explored in TTS, including FastSpeech2s~\cite{ren2020fastspeech}, EATS~\cite{donahue2020end}, VITS~\cite{kim2021conditional,shirahata2023period}, etc., seeking to alleviate the mismatch between acoustic feature generation and waveform reconstruction by integrating these two stages together.

In this paper, inspired by the high audio quality of VITS~\cite{kim2021conditional}, we propose Prosody-aware VITS (PAVITS) for EVC, a novel end-to-end system with implicit prosody modeling to enhance content naturalness and emotional naturalness. To our best knowledge, PAVITS is the first EVC method in solving the mismatch between acoustic feature conversion and waveform reconstruction. Compared to original VITS, our approach involves several key innovations.
In order to improve content naturalness with speech quality, we build upon VITS to solve the two-stage mismatch in EVC, and apply multi-task learning since TTS can significantly reduce the mispronunciation. To enhance emotional naturalness, we introduce an emotion descriptor to capture prosody differences associated with different emotional states in speech. By utilizing Valence-Arousal-Dominance values as condition, emotional representation at utterance-level is learned. Latent code is further refined by a prosody integrator, which incorporates with speaker identity and linguistic content to model finer-grained prosody variations. Then frame-level prosody features are obtained from normalizing flow. We also introduce a prosody predictor that leverages emotion labels and phoneme-level text embedding to predict frame-level emotional prosody features. Finally, we devise a prosody alignment loss to connect two modalities, aligning prosody features obtained from audio and text, respectively.

\section{Proposed method}
\label{sec:format}

As shown in Figure 1, inspired by VITS~\cite{kim2021conditional}, the proposed model is constructed based on conditional variational autoencoder (CVAE), consisting of four parts: a textual prosody prediction module, an acoustic prosody modeling module, an information alignment module, and an emotional speech synthesis module. 

The textual prosody prediction (TPP) module predicts the prior distribution $\mathcal\ p\left(z_1 \mid c_1\right)$ as:
\begin{equation}
	\begin{aligned}
			z_1=TPP\left(c_1\right) \sim p\left(z_1 \mid c_1\right) 
	\end{aligned}
	\label{eq1}
\end{equation}
where $\mathcal\ c_1$ including text $\mathcal\ t$ and emotion label $\mathcal\ e$.

The acoustic prosody modeling (APM) module disentangles emotional features with intricate prosody variation, speaker identity, and linguistic content from the source audio given emotion label, forming the posterior distribution $\mathcal\ q\left(z_2 \mid c_2\right)$ as: 
\begin{equation}
	\begin{aligned}
	z_2=APM\left(c_2\right) \sim q\left(z_2 \mid c_2\right) 
	\end{aligned}
	\label{eq2}
\end{equation}
where $\mathcal\ c_2$ including audio $\mathcal\ y$ and emotion label $\mathcal\ e$.

The information alignment module facilitates the alignment of text and speech, as well as the alignment of textual and acoustic prosody representations. In emotional speech synthesis (ESS) module, the decoder reconstructs waveform  $\mathcal\ \hat{y}$ according to latent representation  $\mathcal\ z$. 
\begin{equation}
	\begin{aligned}
	\hat{y}=Decoder\left(z\right) \sim p\left(y\mid z\right)
	\end{aligned}
	\label{eq3}
\end{equation}
where $\mathcal\ z$ comes from $\mathcal\ z_1$ or $\mathcal\ z_2$.

While the proposed model can perform both EVC and emotional TTS after training, EVC will be the main focus of this paper. In the following, we will introduce the details of the four modules.
\begin{figure}[t]
	\centering
	\includegraphics[width=\columnwidth]{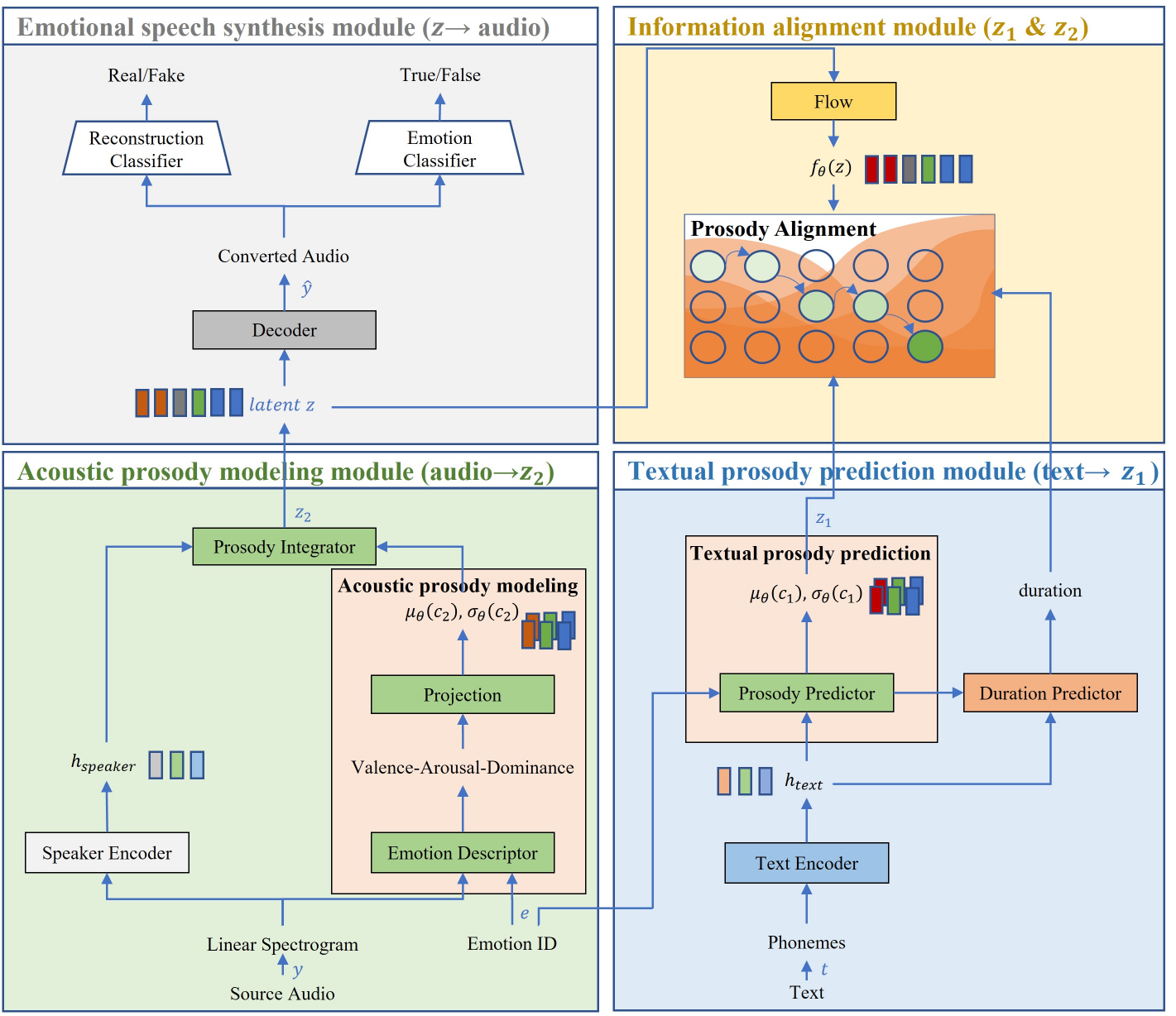}
	\caption{Architecture of PAVITS.}
	\label{Fig1}
\end{figure}

\subsection{Textual prosody prediction module}
Given condition $\mathcal\ c_1$ including text $\mathcal\ t$ and emotion label $\mathcal\ e$, the textual prosody prediction module provides the prior distribution $\mathcal\ p\left(z_1 \mid c_1\right)$ of CVAE. The text encoder takes phonemes as input and extracts linguistic information $\mathcal\ h_{text}$ at first. Considering the extensive prosody variation associated with each phoneme, we employ a prosody predictor to extend the representation to frame-level and predict the prosody variation (a fine-grained prior normal distribution with mean $\mathcal\ \mu_\theta$ and variance $\mathcal\ \sigma_\theta$ generated by a normalizing flow $\mathcal\ f_\theta$) based on emotion label.  
\begin{equation}
	p\left(z_1 \mid c_1\right) = N\left(f_\theta\left(z_1\right) ; \mu_\theta\left(c_1\right) ; \sigma_\theta\left(c_1\right)\right)\left|\operatorname{det} \frac{\partial f_\theta\left(z_1\right)}{\partial z}\right| \\ 
	\label{eq4}
\end{equation}

\textit{Text Encoder}: Since the training process is constrained by the volume of textual content within parallel datasets, we initially convert text or characters into a phoneme sequence as a preprocessing step to maximize the utility of the available data,  resulting in improved compatibility with the acoustic prosody modeling module. Similar to VITS~\cite{kim2021conditional}, text encoder comprises multiple Feed-Forward Transformer (FFT) blocks with a linear projection layer for representing linguistic information.

\textit{Prosody Predictor}: Prosody predictor leverages phoneme-level linguistic information extracted by the text encoder to anticipate frame-level prosody variation given discrete emotion label. It has been observed that simply increasing the depth of stacked flow does not yield satisfactory emotional prosody variations, unlike the prosody predictor. Therefore, the inclusion of the prosody predictor guarantees a continuous enhancement in prosody modeling for both the TPP and APM modules. The prosody predictor comprises multiple one-dimensional convolution layers and a linear projection layer. Furthermore, we integrate predicted emotional prosody information with linguistic information as input for the duration predictor, which significantly benefits the modeling of emotional speech duration.

\subsection{Acoustic prosody modeling module}
The acoustic prosody modeling module provides emotional features with fine-grained prosody variation based on dimensional emotion representation, i.e., Valence-Arousal-Dominance values. Speaker identity and speech content information are also disentangled from the source audio and then complete feature fusion through the prosody integrator as the posterior distribution $\mathcal\ q\left(z_2 \mid c_2\right)$.
\begin{equation}
		q\left(z_2 \mid c_2\right) = N\left(f_\theta\left(z_2\right) ; \mu_\theta\left(c_2\right) ; \sigma_\theta\left(c_2\right)\right) \\ 
	\label{eq5}
\end{equation}

\textit{Speaker encoder}: Considering the APM module's increased focus on understanding emotional prosody more thoroughly compared to previous models, it's apparent that speaker characteristics could unintentionally be overlooked during conversion. Recognizing the critical role of fundamental frequency (F0) in speaker modeling~\cite{busso2009analysis}, we augment the F0 predictor of~\cite{zhang2022visinger} by adding multiple one-dimensional convolutional layers and a linear layer to construct the speaker encoder, which tackles the issue effectively.

\textit{Emotion descriptor}: To enhance PAVITS's emotional naturalness, we employ a specific SER system rooted in Russell's circumplex theory~\cite{russell1980circumplex} to predict dimensional emotion representation, encompassing Valence-Arousal-Dominance values as a conditional input. This input guides the capture of nuanced prosody variations, which ensures that while satisfying human perception of emotions at utterance-level, natural prosody variations are retained from segment-level down to frame-level, preserving intricate details. It consists of a SER module~\cite{wagner2023dawn} and a linear projection layer.  

\textit{Prosody Integrator}: The prosody integrator incorporates a combination of speaker identity attributes, emotional prosody characteristics, and intrinsic content properties extracted from the linear spectrogram. It is constructed using multiple convolution layers, Wavenet residual blocks, and a linear projection layer.

\subsection{Information alignment module}
In VITS~\cite{kim2021conditional}, the existing alignment mechanism, which is called Monotonic Alignment Search (MAS), solely relies on textual and acoustic features from parallel datasets. Thus, it is insufficient in capturing emotional prosody nuances, hindering effective linkage between the TPP and APM modules. To overcome this limitation, we propose an additional prosody alignment loss function based on Kullback-Leibler divergence, to facilitate joint training for frame-level prosody modeling across the TPP and APM modules, with the goal of enhancing prosody information integration and synchronization within our model.
\begin{equation}
	L_{psd} = D_{KL}\left(q\left(z_2 \mid c_2\right)\|\ p\left(z_1 \mid c_1\right)\right) \\
	\label{eq6}
\end{equation}

\subsection{Emotional speech synthesis module}
In the emotional speech synthesis module, the decoder generates a waveform based on latent $\mathcal\ z$, employing adversarial learning to continuously enhance naturalness in both content and emotion. 
To improve the naturalness of content, $\mathcal\ L_{\text {recon\_cls }}$ minimizes the L1 distance between predicted and target spectrograms, $\mathcal\ L_{\text {recon\_fm }}$ minimizes the L1 distance between feature maps extracted from intermediate layers in each discriminator, aimed at enhancing training stability. Since the former predominantly influences the early-to-mid stage, while the latter assumes a more prominent role in mid-to-late stage, we introduce two coefficients to balance their contributions as follows.
\begin{equation}
	\begin{aligned}
		L_{recon }=\gamma L_{recon\_cls }+\beta L_{recon\_fm }(G)
	\end{aligned}
	\label{eq7}
\end{equation}
To enhance the perception of emotions, $\mathcal\ L_{\text {emo\_cls }}$ represents the loss function for emotional classification, while $\mathcal\ L_{\text {emo\_fm }}$ denotes the loss associated with feature mapping for emotion discrimination.
\begin{equation}
	\begin{aligned}
		L_{emo }=L_{emo\_cls }+ L_{emo\_fm }(G)
	\end{aligned}
	\label{eq8}
\end{equation}

\subsection{Final loss}
By combining CVAE with adversarial training, we formulate the overall loss function as follows:
\begin{equation}
	L=L_{recon }+L_{adv }(G)+L_{emo }+L_{psd }+L_{F0 }+L_{dur }
	\label{eq9}
\end{equation}
\begin{equation}
	L(D)=L_{adv }(D)
	\label{eq10}
\end{equation}
where $L_{adv }(G)$ and $L_{adv }(D)$ represent the adversarial loss for the Generator and Discriminator respectively, $L_{F0}$ minimizes the L2 distance between the predicted F0 and corresponding ground truth, $L_{dur}$ minimizes the L2 distance between the predicted duration and ground truth which is obtained through estimated alignment.

\subsection{Run-time conversion}
At runtime, there are two converting methods: a fixed-length approach (Audio-$z_2$-Audio, named PAVITS-FL) and a variable-length approach (Audio-Text-$z_1$-Audio, named PAVITS-VL). The former uses APM module for latent $z$ prediction from audio, ensuring robustness as it remains unaffected by text encoding, but is constrained by a fixed spectrum length due to Dynamic Time Warping (DTW) limitations. The latter employs TPP module to predict latent $z$ from corresponding text obtained through automatic speech recognition (ASR) techinique, which is not bound by duration modeling and offers greater naturalness. Finally, the ESS module's decoder takes latent $z$ (either $z_1$ or $z_2$) as input and synthesizes the converted waveform without a separate vocoder.

\section{Experiments}
\label{sec:pagestyle}

\subsection{Dataset}
We perform emotional conversion on a Mandarin corpus belonged to Emotional Speech Dataset (ESD)~\cite{zhou2022emotional} from neutral to angry, happy, sad, and surprise, denoted as \textit{Neu-Ang}, \textit{Neu-Hap}, \textit{Neu-Sad}, \textit{Neu-Sur} respectively. For each emotion pair, we use 300 utterances for training, 30 utterances for evaluation, and 20 utterances for test. The total duration of training data is around \textit{80 minutes} (\textit{16 minutes} per emotion category), which is absolutely small compared to others.

\begin{table}[]
\centering
\small 
\caption{A comparison of MCD [dB] values. }
\resizebox{\linewidth}{!}{
	\begin{tabular}{l|llll}
	\hline
	\multicolumn{1}{c|}{\multirow{2}{*}{Model}} & \multicolumn{4}{c}{MCD {[}dB{]}}                                                                                      \\ \cline{2-5} 
	\multicolumn{1}{c|}{}                       & \multicolumn{1}{c}{Neu-Ang} & \multicolumn{1}{c}{Neu-Hap} & \multicolumn{1}{c}{Neu-Sad} & \multicolumn{1}{c}{Neu-Sur} \\ \hline
	CycleGAN                                    & 4.41                        & 4.24                        & 4.32                        & 5.68                        \\
	StarGAN                                     & 4.52                        & 4.46                        & 4.31                        & 5.79                        \\
	Seq2seq-WA2                                 & 3.73                        & 3.72                        & 3.77                       & 5.60              \\
	VITS                                        & 3.68                        & 3.70                        & 3.69                        & 5.41                        \\ \hline
	PAVITS-FL (proposed)                                    & \textbf{3.42}               & 3.63                        & 3.40                        & 4.61                        \\
	PAVITS-VL (proposed)                                    & 3.58                        & \textbf{3.62}               & \textbf{2.98}               & \textbf{3.96}                        \\ \hline
\end{tabular}
}
\end{table}

\subsection{Experimental Setup}
We train the following models for comparison. 
\begin{itemize}
	\item CycleGAN~\cite{kaneko2019cycleganvc2} (\textit{baseline}): CycleGAN-based EVC model with WORLD vocoder.
	\item StarGAN~\cite{choi2020stargan} (\textit{baseline}): StarGAN-based EVC model with WORLD vocoder.
	\item Seq2seq-WA2~\cite{zhou2021limited} (\textit{baseline}): Seq2seq-based EVC model employing 2-stage training strategy with WaveRNN vocoder. 
	\item VITS~\cite{kim2021conditional} (\textit{baseline}): EVC model constructed by original VITS, operating independently in both fixed-length and variable-length, take the average as the result.
	\item PAVITS-FL (\textit{proposed}): the proposed model based on VITS, incorporates all the contributions outlined in the paper, but operate within a fixed-length framework.
	\item PAVITS-VL (\textit{proposed}): the proposed model based on VITS, incorporates all the contributions outlined in the paper, but operate within a variable-length framework leveraging ASR to obtain text from source audio.
\end{itemize}

\begin{table*}[t]
\centering
\small 
\renewcommand{\arraystretch}{1.0}
\caption{Experimental results in terms of subjective mean opinion score (MOS) }
	\begin{tabular}{c|cccccccc}
		\hline
		\multirow{3}{*}{EVC Model} & \multicolumn{8}{c}{MOS}                                                                                                                                                                    \\ \cline{2-9} 
		& \multicolumn{4}{c|}{Speech Quality}                                                                    & \multicolumn{4}{c}{Naturalness}                             \\ \cline{2-9} 
		& Neu-Ang            & Neu-Hap            & Neu-Sad            & \multicolumn{1}{c|}{Neu-Sur}            & Neu-Ang            & Neu-Hap            & Neu-Sad            & Neu-Sur            \\ \hline
		CycleGAN                   & 3.91±0.19          & 4.04±0.16          & 3.95±0.13          & \multicolumn{1}{c|}{3.84±0.12}          & 3.83±0.19          & 4.01±0.21          & 3.86±0.20          & 3.90±0.14          \\
		StarGAN                    & 3.53±0.10          & 3.50±0.12          & 3.46±0.14          & \multicolumn{1}{c|}{3.49±0.07}          & 3.56±0.20          & 3.61±0.14          & 3.71±0.18          & 3.70±0.17          \\
		Seq2seq-WA2               & 3.95±0.14          & 4.03±0.24          & 4.14±0.29          & \multicolumn{1}{c|}{4.03±0.16}          & 3.72±0.14          & 3.67±0.15          & 3.72±0.17          & 3.89±0.20          \\
		VITS                       & 4.49±0.06          & 4.40±0.13          & 4.55±0.12          & \multicolumn{1}{c|}{4.51±0.06}          & 4.00±0.19          & 4.15±0.12          & 4.23±0.20          & 4.26±0.15          \\ \hline
		PAVITS-FL (proposed)                   & 4.62±0.04          & 4.62±0.04          & \textbf{4.64±0.04} & \multicolumn{1}{c|}{\textbf{4.66±0.02}} & 4.25±0.19          & 4.44±0.09          & 4.48±0.07          & 4.40±0.13          \\
		PAVITS-VL (proposed)                   & \textbf{4.72±0.02} & \textbf{4.72±0.01} & 4.63±0.03          & \multicolumn{1}{c|}{4.66±0.03}          & \textbf{4.39±0.14} & \textbf{4.60±0.11} & \textbf{4.59±0.05} & \textbf{4.61±0.10} \\ \hline
		Ground Truth               & 4.78±0.02          & 4.81±0.01          & 4.82±0.01          & \multicolumn{1}{c|}{4.86±0.01}          & 4.71±0.06          & 4.78±0.05          & 4.83±0.02          & 4.80±0.04          \\ \hline
	\end{tabular}
\end{table*}

\begin{table}[]
	\centering
	\small 
	\caption{Ablation study with MOS test}
	\makebox[0.1\textwidth][c]{
		\begin{tabular}{lccc}
			\hline
			\multicolumn{1}{c}{EVC Model}        & Speech Quality        & Naturalness             \\ \hline
			\multicolumn{1}{c}{PAVITS (proposed)} & \textbf{4.67±0.04} & \textbf{4.60±0.07}  \\
			w/o Prosody Predictor                  & 4.48±0.10 & 4.16±0.13  \\
			w/o Prosody Alignment                  & 4.38±0.05 & 4.08±0.10 \\
			w/o Prosody Integrator                & 4.56±0.09 & 4.37±0.17 \\ \hline
		\end{tabular}
	}
\end{table}

\subsection{Results \& Discussion}
Mel-cepstral distortion (MCD) was calculated for objective evaluation, as depicted in Table 1. In terms of subjective evaluation, Mean Opinion Score (MOS) tests were conducted to appraise both the quality and naturalness of speech as shown in Table 2. The naturalness score was derived by averaging the scores for content naturalness and emotional prosody naturalness, as rated by 24 participants, each of whom assessed a total of 148 utterances. We further report emotional similarity results between converted audio and human voice to gauge emotional naturalness as illustrated in Figure 2.

Through the above-mentioned metrics, it is obvious that the proposed PAVITS achieves competitive performance on both objective and subjective evaluation. From the perspective of objective MCD and subjective MOS, both original VITS and our proposed PAVITS models always outperform other models with traditional vocoder or neural vocoder, which proves that the integration of neural acoustic converter and vocoder is suitable for EVC task to enhance speech quality and naturalness. It is worth noting that even in the case of the fixed-length PAVITS-FL model, there is a reduction of over 0.4 in MCD when compared to the variable-length seq2seq model and the original VITS model. Furthermore, there has been an enhancement of 0.6 and 0.2 in MOS, respectively. To some extent, it reflects how human tend to be influenced by audio quality when assessing model naturalness, especially when there are significant differences in quality being compared.

As depicted in Figure 2, our proposed PAVITS-VL (variable-length) model aligns more closely with human perception in the converted audio, which attributed to the model's capacity for fine-grained granularity in modeling speech emotion, incorporating implicit prosody cues. To further show the effectiveness of our method, we visualize the spectrogram of testing clips, as exemplified in Figure 3. It is readily apparent that the spectrogram converted by PAVITS exhibits finer details in prosody variations within the pertinent frequency bands, while simultaneously preserving descriptive information for other frequency bands. Consequently, the audio generated by PAVITS possesses a prosody naturalness and emotional accuracy that closely approximates the ground truth spectrogram.

\begin{figure}[htbp]
	\centering
	\scalebox{0.95}[0.90]{
	\includegraphics[width=\columnwidth]{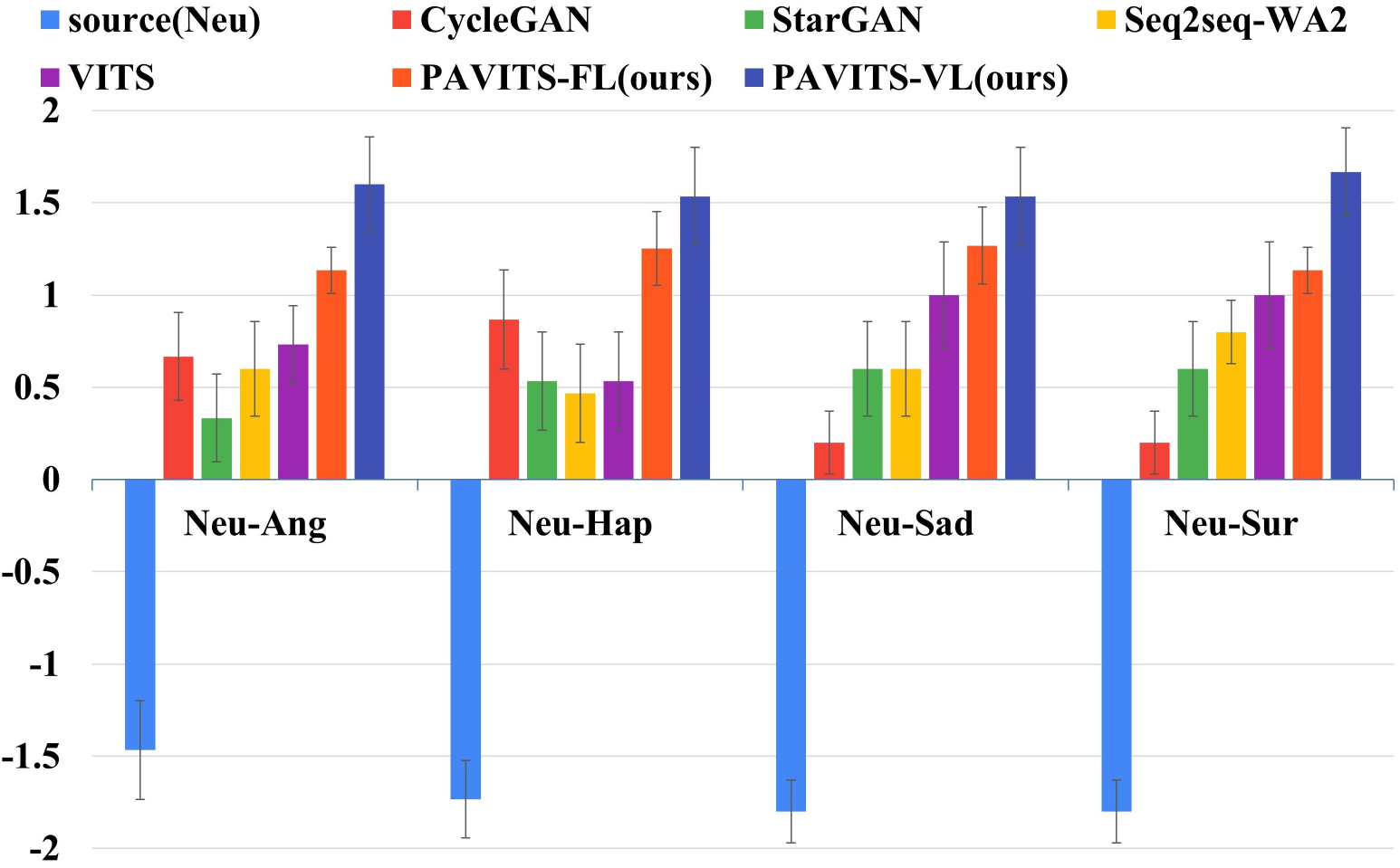}
	}
	\caption{Emotional similarity test with 95\% confidence interval following~\cite{zhou2021limited}.}
	\label{Fig2}
\end{figure}

\begin{figure}[htbp]
	\centering
	\scalebox{0.95}[0.81]{
		\includegraphics[width=\columnwidth]{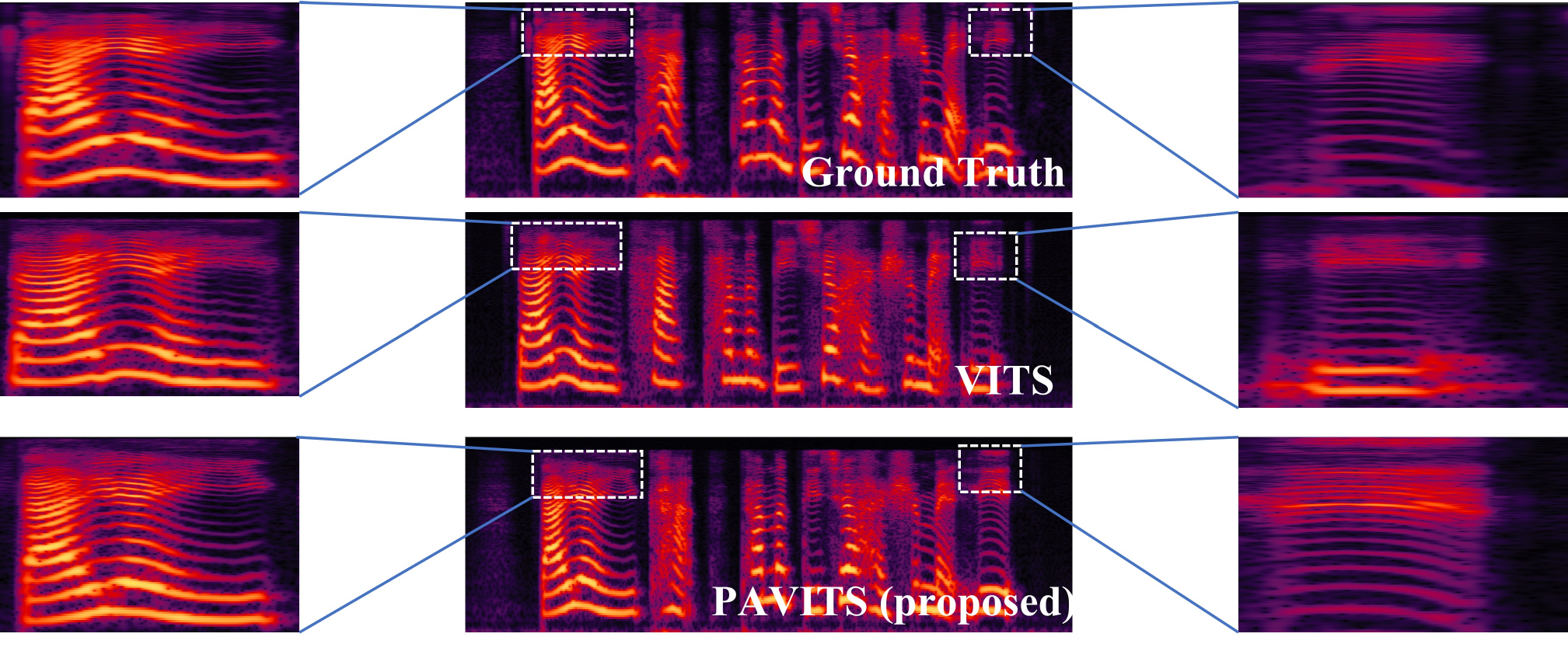}
	}
	\caption{Spectrogram of a testing clip (happy), from top to bottom are ground truth, converted by original VITS, and proposed PAVITS.}
	\label{Fig3}
\end{figure}

\subsection{Ablation Study}
We further conduct an ablation study to validate different contributions. We remove prosody predictor, prosody alignment, and prosody integrator in turn and let the subjects evaluate quality and naturalness of converted audio. From Table 3, we can see that all scores are degraded with the removal of different components.
When remove prosody predictor, the speech quality does not undergo significant changes, as the original VITS primarily relies on textual features as input. However, a significant decrease in naturalness is observed, attributed to the loss of explicit emotion label for TPP module as a conditioning factor. This highlights the importance of aligning with APM module on the basis of information asymmetry, which reflects the ingenious design of prosody modeling structure.
Note that the performance of PAVITS is worse than VITS after deleting prosody alignment, it might be attributed the fact that latent prosody representations are not constrained during training, which damages the original MAS mechanism present in VITS. 
To further show the contribution from the prosody integrator, we replace it with a simple concatenation. Both speech quality and naturalness show a slight decrease, indicating that utilizing prosody integrator for information fusion is quite effective for APM module.

\section{Conclusion}
\label{sec:typestyle}

In this paper, we propose Prosody-aware VITS (PAVITS) for emotional voice conversion (EVC). By integrating acoustic prosody modeling (APM) module with textual prosody prediction (TPP) module through prosody alignment, the fine-grained emotional prosody features across various scales of emotional speech can be learned effectively. Experimental results on ESD corpus demonstrate the superiority of our proposed PAVITS for content naturalness and emotional naturalness, even when dealing with limited data scenarios. In the future, we will explore the controllable emotional prosody modeling to allow better interpretability of EVC.



\small
\bibliographystyle{IEEEbib}
\small
\bibliography{Template}

\end{document}